# Methodology and feasibility of neurofeedback to improve visual attention to letters in mild Alzheimer's disease

Deirdre E. **McLaughlin**[1], Daniel **Klee**[2], Tab **Memmott**[1,2], Betts **Peters**[1], Jack **Wiedrick**[3], Melanie **Fried-Oken**[1,2,4,5], Barry **Oken**[2,4,6*] and the Consortium for Accessible Multimodal Brain-Body Interfaces (CAMBI)

[1] Institute on Development & Disability, Department of Pediatrics, Oregon Health & Science University, Portland, OR, USA

[2] Department of Neurology, Oregon Health & Science University, Portland, OR, USA

[3] Biostatistics & Design Program, Oregon Health & Science University-Portland State University School of Public Health, Portland, OR, USA

[4] Department of Biomedical Engineering, Oregon Health & Science University, Portland, OR, USA

[5] Department of Otolaryngology, Oregon Health & Science University, Portland, OR, USA

[6] Department of Behavioral Neuroscience, Oregon Health & Science University, Portland, OR, USA

**Running Title:** Neurofeedback and visual attention in AD

**Correspondence Author & Address**
Dr. Barry S. Oken M.D., PhD.
Department of Neurology
3250 Southwest Sam Jackson Park Road
Portland, OR 97239 USA
Phone: 503-494-8873
Email: oken@ohsu.edu

**Author Emails**
Deirdre McLaughlin – deirdre@ohsu.edu
Daniel Klee – klee@ohsu.edu
Tab Memmott – memmott@ohsu.edu
Betts Peters – petersbe@ohsu.edu
Jack Wiedrick – wiedrick@ohsu.edu
Melanie Fried-Oken – friedm@ohsu.edu
Barry Oken - oken@ohsu.edu

# ABSTRACT

*Background:* Brain-computer interface (BCI) systems are controlled by users through neurophysiological input for a variety of applications, including communication, environmental control, motor rehabilitation, and cognitive training. Although individuals with severe speech and physical impairment are the primary users of this technology, BCIs have emerged as a potential tool for broader populations, especially with regards to delivering cognitive training/interventions with neurofeedback.

*Methods:* The goal of this study was to investigate the feasibility of using a BCI system with neurofeedback as an intervention for people with mild Alzheimer's disease (AD). The study focused on visual attention and language since AD is often associated with functional impairments in language and reading. This study enrolled 5 adults with mild AD in a nine- to 13-week BCI EEG-based neurofeedback (NFB) intervention to improve attention and reading skills. Two participants completed intervention entirely. The remaining three participants could not complete the intervention phase because of COVID-19 restrictions. Pre- and post-assessment measures were used to assess reliability of outcome measures and generalization of treatment to functional reading, processing speed, attention, and working memory skills.

*Results:* Participants demonstrated steady improvement in most cognitive measures across experimental phases, although there was not a significant effect of NFB on most measures of attention. One subject demonstrated statistically significant improvement in letter cancellation during NFB. All participants with mild AD learned to operate a BCI system with training.

*Conclusions:* Results have broad implications for the design and use of BCI systems for participants with cognitive impairment. Preliminary evidence justifies implementing NFB-based cognitive intervention in AD

*Trial Registration:* This study was registered with ClinicalTrials.gov (NCT03790774).

## INTRODUCTION

Language deficits, including impairments in comprehension and reading, are often present in early Alzheimer's disease (AD) (Storandt et al., 1995). Given the complex relationship between language and other cognitive domains, difficulties with language comprehension and expression affect many areas of daily living and are important contributors to social exclusion (Klimova et al., 2015). Impairments in reading comprehension affect many functional activities, such as taking in news from a newspaper, understanding a book or email, or even working with a computer. Reading difficulties are strongly related to attention and executive function deficits (Kendeou et al., 2014). Since attention is one of the earliest non-memory domains affected in AD (Perry & Hodges, 1999) interventions targeting attention may improve functional areas, such as reading, in people with mild AD.

There is a breadth of behavioral interventions for improving cognitive function both in older adults without known disabilities and in adults with AD. Reviews of intervention practices for cognition in older adults have usually demonstrated domain-specific improvements (Ball et al., 2002; Rolle et al., 2017; Willis et al., 2006) such that training in one cognitive domain (e.g., working memory) does not transfer to untrained domains (e.g., episodic memory) (Sala et al., 2018). According to two systematic reviews, training that targets multiple cognitive domains for people with AD has mixed results, with some studies demonstrating improvements and others showing no functional change (Carrion et al., 2018; Kallio et al., 2017). With advancements in technology, computer-based cognitive training approaches that feature interactive components for participants (e.g., video-gaming elements) have been proposed as a possible intervention modality (Anguera et al., 2013; Barnes et al., 2009; Toril et al., 2014). Historically, cognitive interventions train and measure behavioral responses as indices of cognition. A major limitation of current behavioral interventions is the lack of real-time neurophysiologic metrics to objectively measure and guide the user's learning.

Brain-computer interface (BCI) is a technology that has been developed to enhance, restore or replace physical or cognitive functioning using real-time invasive or noninvasive recording of brain signals as an input method to control the technology (Wolpaw & Wolpaw, 2013). For example, the sensorimotor rhythm has been used in BCIs for control of computers, robotic arms, and wheelchairs by individuals with tetraplegia (Wolpaw & Wolpaw, 2013). While much BCI research has involved healthy adults, a major clinical benefit of BCI is for people with disabilities (Kübler et al., 2006) who may directly benefit from BCIs designed to restore or replace physical function. For individuals with severe speech and physical impairments secondary to incomplete locked-in syndrome, spelling BCIs have been designed to use the P300 signal, an endogenous event-related potential following a salient stimulus, as a 'keystroke' or intended selection for communication (Akcakaya et al., 2014; Rezeika et al., 2018). In this paradigm, target letter presentations are interspersed with non-target presentations. Perception of the salient target results in an attentional neurophysiological event-related potential (P300) to the intended target. This task requires the user to employ many cognitive skills, including (but not limited to) sustained visual attention to the screen, remembering the target throughout the task (working memory), selective attention to the targets on the screen amidst non-targets, and divided attention between the presented letters and feedback. As such, training with a P300 spelling paradigm may be beneficial across multiple cognitive domains.

Neurofeedback (NFB) is the delivery of real-time electroencephalography (EEG) based feedback. Few clinical NFB present cognitive tasks; most simply provide feedback to ongoing EEG frequencies such as alpha or theta band power (Evans et al., 2020; Sitaram et al., 2017). Some NFB studies have used very simple recordings, e.g., 2-channel forehead strap systems that may be providing feedback based on the frequency of scalp muscle activity rather than brain activity (Paluch et al., 2017). Other NFB systems may be giving NFB based on simple alertness (Oken et al., 2018) which is known to affect ERPs (Huang et al., 2017) and BCI performance (Oken et al., 2018). To improve cognitive performance, we believe use of NFB during performance of a cognitive task is better than using a simple rest state and thus there has been some merging of NFB and BCI research. This merging of BCI and NFB is also relevant, since training people to optimally use a BCI system may require NFB during BCI use.

While the P300 may be used to guide the BCI's decision-making process for inferring the user's intention, most P300 speller systems require too many stimulus presentations in order to generate the P300, and thus cannot not provide timely NFB. One group has used presentation of the chosen letter during use of a P300 speller as feedback, that improve performance at high letter flash rates, but this behavioral feedback is not quite the same as NFB (Arvaneh et al., 2015). The same group followed up with similar correct letter feedback and demonstrated changes in alpha activity at Pz consistent with greater attention during feedback training compared to no-feedback training (Arvaneh et al., 2019). As such, utilizing some EEG band measure activity may be a way to improve attention to BCI cognitive demands. This study used alpha activity for NFB since it has been used previously for NFB during tasks (Jirayucharoensak et al., 2019; Lavy et al., 2019) and we had some preliminary data (see below in Calculation of Neurofeedback section and Figure 3). Training with EEG-based NFB to enhance attention is an important research direction in cognitive training (Ordikhani-Seyedlar et al., 2016). There is some evidence of benefits of this training delivery method, including increased visual attention in young adults without disabilities (Arvaneh et al., 2019) and improvement of attention deficit hyperactivity disorder symptoms in children (Lim et al., 2012). In terms of aging, most studies based on recent systematic reviews have focused on healthy older adults (Jiang et al., 2017; Laborda-Sánchez & Cansino, 2021). One randomized wait-list controlled study of adults without disabilities (n = 39), including 24 half-hour sessions of BCI training using a Stroop task, noted slight improvements on the Repeatable Assessment of Neuropsychological Status (Lee et al., 2015). A larger follow-up study by the same group in 240 health elderly did not demonstrate clear improvements in the same cognitive outcomes but did demonstrate a sex moderation effect that suggested men improved more with the intervention than those men in the waitlist control with female participants showing no significant difference (Yeo et al., 2018). The limited number of studies included in systematic reviews that involved people with cognitive impairments (mild cognitive impairment or AD) all had significant weaknesses; inclusion of NFB as one component of a more general lifestyle intervention (Fotuhi et al., 2016); use of only a single channel of EEG (Luijmes et al., 2016); lack of a control group or multiple baseline assessments to control for learning effects (Jang et al., 2019; Lavy et al., 2019; Surmeli et al., 2016); and use of a simple retrospective analysis of a poorly-characterized group of people of mixed ages with cognitive dysfunction treated with NFB (Koberda, 2014). One higher-quality paper studied 65 women with amnestic mild cognitive impairment and 54 control women who received NFB in a gamified interface (Jirayucharoensak et al., 2019). There were two control groups: a treatment as usual (passive control) and an exergame done as frequently as the NFB sessions (active control). The NFB and the active control group both demonstrated improvements in some, but not the same, cognitive outcomes.

The aim of this feasibility study was to adapt an existing BCI system, BciPy, (Memmott et al., 2021) (available from https://github.com/CAMBI-tech/BciPy), to also provide NFB and explore its application to facilitate improvement in attention and reading skills in people with mild AD. Outcome measures targeted skills within the domain of the proposed intervention, including selective attention to letters, speed of processing while reading, and working memory while manipulating letters. We hypothesized that NFB training would improve attention to letter processing and other attention-dependent measures related to functional reading. This paper reports on the methodology of a small feasibility study demonstrating proof of concept for this novel NFB training method for individuals with mild AD

**MATERIALS AND METHODS**

*Participants*

Six individuals with possible or probable AD were recruited as study participants. Five participants were enrolled; this was the target number for this feasibility study. One participant was not enrolled because of study visit demands. The participant number for this US National Institutes of Health-funded study was small but is typical for the planned analysis: a non-experimental multiple baseline single case research design (SCRD) (Gabler et al., 2011; Ledford & Gast, 2018). Ultimately a conventional randomized controlled trial will be best for demonstrating efficacy of a NFB intervention, given concerns ranging from financial issues to placebo and expectancy effects (Oken, 2008; Thibault et al., 2018). However, the high cost of this particular NFB intervention, requiring many visits to participants' homes or other preferred locations, makes the SCRD ideal for this small feasibility study. Due to safety concerns involved in conducting home visits for data collection with vulnerable populations during the COVID-19 pandemic, the study was concluded with complete data from only two participants. These participants, #1 and #2, will be referred to as "completers". Data collection for the other three participants was discontinued before or shortly after the beginning of the intervention phase. Please refer to Table 1 for participant description and demographic information.

Participants were recruited from the Oregon Health & Science University (OHSU) Layton Aging and Alzheimer's Center, which is funded in part as an NIH Alzheimer's Disease Research Center. Recruitment procedures included chart review, communication with the treating neurologist regarding eligibility criteria for diagnosis of pAD, and a phone screening that consisted of approximately 30 minutes of questioning to determine eligibility. During this phone screening, participants completed the judgment subtest of the Neurobehavioral Cognitive Status Examination (NCSE) in order to assess capacity to consent to the study procedures (Kiernan et al., 1987). Inclusion criteria were as follows: (1) adults aged 50-100 years old; (2) a diagnosis of mild possible or probable AD by a medical diagnosis by their treating neurologist at the Layton Aging and Alzheimer Disease Center consistent with conventional criteria (McKhann et al., 1984) (3) a Global Clinical Dementia Rating (CDR) of 0.5 or 1.0 (Morris, 1993), (4) a Montreal Cognitive Assessment (MoCA) score of $\geq 14$ (Nasreddine et al., 2005) or a Mini Mental State Examination (MMSE) $\geq 18$ (Upton, 2013) within the previous two months; (5) a mild language impairment attributable to cognitive challenges assessed through clinical impression by either an ASHA-certified speech-language pathologist from the research team (DEM) or behavioral neurologist at the Layton Center; (6) absence of EEG-altering medications (e.g. benzodiazepines); (7) no significant motor, vision, or hearing impairment; and (8) greater than 80% accuracy on at least one of four trials for a practice RSVP task (described below). This study was approved and overseen by the OHSU Institutional Review Board (IRB #18984) and registered with ClinicalTrials.gov (NCT03790774). All methods were performed in accordance with the relevant guidelines and regulations.

To ensure compliance with the study procedures and facilitate communication and scheduling of study activities, participants were required to enroll with a study partner who also consented to participation in the study. Both participants and study partners provided written informed consent. Study participants were paid at each visit, with a completion bonus paid after follow-up; study partners were not paid for their participation. In order to qualify as a study partner, an indivdidual needed to: (1) have a close relationship with the participant (e.g. spouse, adult child, or friend); (2) interact an average of at least 5 hours per week with the study participant, either in person or over the phone; (3) demonstrate capacity to consent as evidenced by a score of ≥3 on the NCSE Judgement subtest; and (4) demonstrate low risk of cognitive impairment, as measured by a score of ≥31 on the modified Telephone Interview for Cognitive Status (TICS-m) (Knopman et al., 2010). The study partners consisted of one brother, one son, one daughter, one mother, and one spouse.

*Procedure*

This study followed a within-subject A-B design with multiple data collection sessions both before and during administration of the intervention (hereafter "baseline" and "intervention", respectively). In the baseline phase, participants received RSVP Keyboard training without NFB. In the intervention phase, the training incorporated NFB. A single follow-up session was conducted 4-5 weeks after the end of the intervention phase. All sessions were conducted at either the participant's home, the Oken Laboratory at OHSU, or a neutral location, e.g., a public library meeting room, according to the participant's preference. Consistent days of the week, start times, and visit durations were maintained for each participant. Please see figure 1 for an outline of study phases and the activities and assessments conducted at each visit; these are described below.

Study Entry Visit and Outcome Measures

The first visit included informed consent procedures, study eligibility questions, and administration of initial summative outcome measures. The summative outcome measures, given once prior to the baseline sessions and once at the final follow-up session, were: (1) the Discourse Comprehension Test (DCT) (Brookshire & Nicholas, 1993), a measure of listening comprehension; and (2) the forward and backward digit span subtests of the Wechsler Adult Intelligence Scale IV (WAIS-IV), a measure of phonological working memory and attention (Wechsler, 2008). The Discourse Comprehension Test requires a participant to listen to a 150 to 200 word short story and respond to comprehension questions (Brookshire & Nicholas, 1993). There are five short stories per test form with eight questions about each story (Brookshire & Nicholas, 1993). The digit span subtests in the WAIS-IV require a participant to listen to a sequence of numbers read aloud and repeat that sequence back to the examiner in the same order (forward) or in reverse order (backward) (Wechsler, 2008). The number of digits in each sequence ranges from two to 16 (Wechsler, 2008). Participants answered questions about their health, demographics, and the nature of their relationship with their study partner. Additionally, participants completed a computer-based practice RSVP Keyboard task without EEG set up or input; this was designed to familiarize them with the demands of the experimental task and to confirm that they could demonstrate the requisite skills, including: (1) attending to targets; (2) responding if a target was present on screen; and (3) inhibiting responses to non-targets.

Participants were presented with a target letter and asked to attend to the letters in a series of 10 letters in order to answer the question: "Was the target letter in the sequence?" Five out of 10 sequences contained the target and five did not. Participants were trained from slowest presentation speed (1 Hz) to fastest presentation speed (4 Hz) in four steps (i.e., 1, 2, 3, 4 Hz). At each step, participants were trained to criterion, defined as getting eight out of 10 items correct at a given presentation rate (1 Hz, 2 Hz, 3 Hz, 4 Hz). Participants were provided up to four chances to train to criterion at a given presentation rate. An exclusion criteria of study enrollment was not achieving 80% accuracy at the 4 Hz presentation rate. No participants were excluded for this reason. This practice RSVP task was repeated at the beginning of each baseline, intervention, and follow up session in order to ensure that participants maintained an ability to complete the task with a presentation rate of 4 Hz.

Baseline Visits and Outcome Measures

Baseline sessions were planned to begin one week after the initial visit, and to occur once per week afterwards. Participants completed four to seven baseline sessions, until there was reasonably stable performance of the outcome measures as assessed by visual analysis (Gast & Spriggs, 2014). During each baseline session, participants completed RSVP Keyboard calibration and copy-spelling tasks (Oken et al., 2014) using the BciPy software (Memmott et al., 2021), as well as repeated measures outcome tasks to monitor progress. The repeated measures tasks were: (1) letter cancellation task; (2) letter span task; and (3) Woodcock Johnson Test of Achievement 4th edition (WJTA-IV) Sentence Reading Fluency Subtest (form A, B, or C) (Mather & Wendling, 2015). A description of each task follows. Participants were not excluded due to performance on these metrics.

For all RSVP Keyboard tasks (with exception to practice task), participants wore a dry electrode cap (Wearable Sensing; San Diego, CA) that measured EEG responses to target and non-target letters (see section on Electrophysiological Recordings and Processing below for more details on EEG recording).

In the BCIpy calibration task, the participants were shown a single target letter and asked to attend to a rapidly-presented series of letters (nine non-target and one target), looking for the target letter. Each letter was presented centrally on the screen, one at a time, at a rate of 3 Hz. The temporal position of the target letter in each sequence was randomly assigned. For each of 100 sequences of letters, participants mentally responded when they saw the target on the screen. The intention of this paradigm was to elicit a P300 signal in response to target letter presentations, and to gather data for training a classifier to be used in the copy-spelling task. The calibration task lasted approximately 13 minutes. The main outcome measure in calibration was area under the curve (AUC), a measure of classification accuracy. Software and specifics related to the BciPy classifier are more fully outlined in Memmott and colleagues (2021). For classification, the software uses a regularized discriminant analysis with 10-fold cross validation to evaluate EEG in the 500 ms epoch following task stimuli.

In the copy-spelling task, participants were asked to copy a phrase letter by letter ("HELLO_" followed by a word of their choice) by selecting each target letter in the phrase from a stream of non-target letters. Participants were instructed to mentally react to a target letter, as in the calibration task. For each sequence, the system combined EEG evidence with probabilities determined by an integrated language model (Oken et al., 2014) repeating sequences until one letter reached a probability threshold of 0.80 and was selected for typing. This task demands vigilance to the target for several sequences, as the target letter typically must be selected more than once before the system comes to a decision. The main outcome measure for the copy-spelling task were whether or not the target word was successfully copied.

The letter cancellation task in this study was adapted from the letter cancellation task used in Baddeley and colleagues (2001), and measured selective attention. Participants were instructed to cross out all instances of a target letter, "Z", as quickly as possible, in a grid of upper-case letters presented on an standard letter size sheet of paper in 14-point Arial font (Baddeley et al., 2001). There were two forms that varied in difficulty: an easier version with a combination of 10 curved letters as foils (e.g., "B", "P", and "R") and a harder version with a combination of 10 straight letters as foils (e.g., "K", "M", and "Y"). Baddeley and colleagues (2001) found that participants with AD took significantly longer on a version with straight letter foils, and proposed that this effect was attributable to the similarities of visual features between the non-target letters and the target letter "Z" (Baddeley et al., 2001). There were 20 "Z" targets on each form out of a total of 150 letters. Target positions were randomized and plotted by assigning random x and y values to a 10 by 15 grid using R version 3.6.1 (R core team, n.d.). Randomly-generated form versions with more than three adjacent targets in a row, column, or corner were rejected. The outcome measure for this task was total completion time (in seconds) corrected for task accuracy (total time/accuracy), as used in other studies.

The letter span task in this study, which measures working memory, was adapted from the letter span task used by Conrad and Hull (1964) with task considerations modeled from the WAIS digit span subtest (Conrad & Hull, 1964; Wechsler, 2008). Participants attended to a sequence of two to eight letters, presented one at a time at a rate of one per second on a monitor, and recite the sequence back to the examiner either in the same order (forward condition) or in reverse order (backward condition). There were two items for each sequence length in the task. The task was discontinued when the participant answered both items for a given sequence length incorrectly. The letter span task was programmed in Python using PsychoPy3 v3.0.0b11 (Peirce et al., 2019). Strings consisted of only consonant letters to reduce the ability for participants to use a word encoding strategy. Sequences were reviewed by three researchers to remove any consonant combinations commonly used as acronyms or abbreviated phrases in English (e.g., BRB, HQ, RSVP). Stimulus order was randomized and 15 unique versions were generated to reduce the chance of a repetition learning effect across baseline, intervention, and follow-up weeks for each condition. Participants received unique versions in a randomized order. The outcome measure for this task was maximum sequence length, defined as the longest sequence length where a participant recited at least one of the two sequences correctly.

The well-validated WJTA-IV Sentence Reading Fluency subtest (Mather & Wendling, 2015) was used to measure processing speed. This subtest requires participants to read as many sentences as possible and answer whether the sentence is generally "True" or "False" in three minutes. Example items include: "Fire is hot", "Dogs can eat", and "A school bus has a driver". In order to minimize a repetition learning effect over baseline, intervention, and follow up sessions, the order of the three unique test forms was randomly permuted for each participant (e.g., C, A, B, repeating). The outcome measure for this task was the number of items answered correctly in a three-minute period.

Intervention Visits and Outcome Measures

The decision to begin the intervention phase was determined by observation of reasonably stable baseline performance on the three repeated outcome tasks and the BCI calibration task as assessed by visual analysis (Gast & Spriggs, 2014), or by the participant reaching the pre-determined maximum of seven baseline sessions. Intervention sessions were completed three times per week for a six-week period, and required participant to complete a calibration task with NFB and a copy-spelling task. In this phase, the standard RSVP Keyboard calibration task from baseline was adjusted to feature an additional NFB display after each letter sequence. This display was onscreen for two seconds and included five colored boxes ranging from dark red (poor attentional performance) to dark green (excellent attentional performance; see Figure 2). A thick white border around one box was used to indicate to participants their attention rating on the most recent sequence. Participants were asked to try and achieve as many dark green (excellent) ratings as possible during each session and to pay more attention to the RSVP Keyboard sequences if they were given dark red, orange, yellow, or light green ratings. NFB was individualized for each participant and updated weekly based on data from the previous week's calibration task (see Calculation of Neurofeedback below). To monitor progress, participants completed repeated measures tasks before the calibration and copy-spelling tasks during the third session of each week.

<u>Calculation of Neurofeedback</u>

In order to quantify attention to the RSVP Keyboard display for NFB, posterior parieto-occipital alpha power was used to measure engagement of visual attention. This signal was used for NFB because: (1) event-related alpha attenuation during visual tasks is associated with mental effort (Pfurtscheller & Lopes da Silva, 1999; Thut et al., 2006); (2) alpha power has been used in prior NFB experiments during rest states to improve visual cognition (Schwartz & Andrasik, 2016); and (3) results of a pilot study (below) supported its utility for this purpose.

In the alpha power pilot study, participants without known disabilities (n = 8, age range 21-65 years) performed a one-back task that required them to press a button to indicate whether a target sequence was present within a longer sequence of ten letters. The target sequence was always the letter N followed by a random letter, followed by the letter A. Non-target sequences were included to increase difficulty, and consisted of the letters N and A separated by either two or zero other characters. While performing this task, participants wore a 24-channel wet electrode EEG system (BioSemi, Amsterdam). Average button-press error rate for the task was 10%, although three participants made fewer than two errors and were excluded from these analyses. For the five remaining participants, EEG frequency analysis of the 2.5-second epoch including each ten-letter sequence revealed a 17% increase in posterior rhythm (alpha) amplitude when participants made errors compared to when they made no errors (p=0.013, Figure 3). For the three participants who made no errors on the behavioral task, visual inspection of EEG data revealed that they had the lowest amplitude posterior alpha rhythm of the eight participants, further supporting the relationship between attention and posterior alpha rhythm.

Calibration recordings from all available baseline sessions were reviewed in order to determine which of five candidate electrode sites (Pz, Oz, P4, PO7, or PO8) recorded the greatest amount of resting alpha rhythm without interfering artifact for each participant. Due to poor contact at sites Pz and Oz and excessive electromyography (EMG) artifact at sites Oz, PO7, and PO8 for the first two participants, P4 was selected as the channel of interest for calculating NFB during intervention for the completers, participants #1 and #2. Pz was utilized for NFB for participant #4, who finished six intervention sessions over two weeks.

Both participants who completed the study visits consistently demonstrated peak resting alpha activity at approximately 9.0 Hz in both baseline and intervention sessions, therefore, a target frequency band of 8-10 Hz was designated for NFB. All EEG feedback data were bandpass filtered at 7-20 Hz in order to minimize interference of both EMG and a 6 Hz harmonic related to the 3 Hz steady-state visually evoked potential elicited by the RSVP Keyboard letter stream.

Relative power spectral density (PSD; $\mu V^2/Hz$) was calculated using BciPy's signal decomposition module with Welch's method (Welch, 1967) and defined as the PSD of the target band (8-10 Hz) compared to PSD of the wider band (7-20 Hz). From aggregate baseline data within-participants, the 70th, 45th, 30th, and 15th percentiles of relative PSD were set as the NFB cutoffs for the first week of intervention. Specifically, the delineation of the lowest 30% of relative alpha amplitude responses (70th percentile) was used to demarcate dark green "excellent" feedback; the next lowest 25% of relative alpha PSD values (45th percentile cutoff) marked the range of light green "good" feedback, and the remaining 45% of relative PSD values were equally divided into orange ("medium"), yellow ("poor"), and dark red ("bad") ranges, respectively (see Figure 2). These thresholds were chosen in order to provide a positive bias to the feedback and encourage participant engagement. The feedback was presented after each ten-letter sequence (with letters shown at three per second), so approximately every three seconds. After each week of intervention, these cutoffs were recalculated from the most recent week of calibration data.

Follow Up Visit

Approximately one month after the final intervention session, participants completed one follow-up session to assess maintenance of performance on the repeated measures tasks, as well as a final RSVP Keyboard calibration and copy-spelling session without NFB.

Electrophysiological Recordings and Processing

EEG data were recorded using an adjustable DSI VR300 dry-electrode system (Wearable Sensing; San Diego, CA) with linked-ear references, ground at A1 (left earlobe), and scalp electrodes at 10/20 sites FCz, F7, Pz, P4, PO7, PO8, and Oz. Data were sampled at 300 Hz and digitized at 16 bits. The inclusion of an electrode at F7 was a modification of the standard

VR300 design (the original location was P3), made to allow monitoring of electrooculogram. Though field recordings of EEG are sometimes more susceptible to noise than data collected in a more controlled laboratory setting (e.g. ambient 60 Hz electrical noise from power sources in US; mechanical artifact related to appliances or home-medical equipment etc.), all electrodes were calibrated to be within operating ranges as specified by the manufacturer guidelines. Additionally, experimenters monitored real-time EEG during the tasks (displayed using 2-45 Hz bandpass and a 60 Hz notch filters) in order to identify and minimize high amplitude artifacts and ambient noise.

All recordings were down-sampled to 150 Hz and filtered 2-45 Hz for analysis by the classifier. N200 and P300 potentials at site Pz were later quantified offline in Brain Vision Analyzer (Brain Vision LLC; Morrisville, NC, U.S.A.) in order to explore target discrimination and identification processes, respectively. Offline data were band pass filtered 1-45 Hz with a 60-Hz notch before use of independent component analysis (ICA) to remove eye blinks. These data were segmented –200 ms to 1000 ms relative to target and non-target letters before baseline correction using the 200 ms prior to letter presentation and then subjected to artifact rejection. Epochs were flagged for review if they contained voltage steps > 50 μV/ms, amplitude changes > 125 μV over 50 ms, amplitude values > ±75 μV, or sustained amplitude values < 0.5 μV for longer than 100 ms. N200, and P300 peaks were cursored using semiautomatic peak detection windows of 250-400, and 350-500 ms, respectively. N200 and P300 peaks were quantified for analysis using peak-to-trough, or by measuring the change in voltage from the most recent peak of the opposite polarity.

*Data Analysis*

One analysis component for this non-experimental single case research design, as is typical for all SCRD studies, was graphical analysis (Ledford & Gast, 2018). The graphical analysis is particularly useful for heterogenous populations allowing for iterative improvements. However, the multiple repeated baseline and intervention-phase measurements made possible conventional longitudinal statistical analyses as well. We performed a variety of within-subject and between-subject analyses in order to gain better insight into the reliability and informativeness of the outcome measures in the study population.

For analysis of the outcome measures, we omitted the first baseline measurement for each participant on each task. In nearly every case the first recorded outcome measurement was an atypical value, either sharply higher or lower than the next measurement. We interpret this as an acclimation effect of no significance for the study. The first recorded values for outcomes are shown on plots for completeness. For baseline stability and correlation analyses, all baseline points were included.

To evaluate the stability of the measures of interest, we calculated within-subject coefficients of variation (CVs) and intraclass correlation coefficients (ICCs) using the baseline measurements only. The CV is a metric of overall volatility in a longitudinal setting, expressing the equilibrium standard deviation as a fraction of the longrun mean for the participant. Lower CVs are better, and, as a rule of thumb, values of ~5% or lower are indicative of good stability and values in excess of 10% are indicative of poor stability. We estimated an average within-subject standard deviation as the root-mean-square error from an absorbing regression (J. Lloyd Blackwell, 2005) on the baseline measurements of the cohort (absorbing participant effects), and calculated the average within-subject CV as this value divided by the overall baseline mean. (Note that this is a population-level estimate, not a strict average of the individual CVs.) The standard error of the CV was approximated using the formula $\frac{CV}{\sqrt{(2 \times df)}}$, where $df = \frac{(n \times m)}{(1+(m-1)\times r)}$, for n=5 (the number of participants), m=4.4 (the average number of included baseline measurements per participant), and r (an estimate of the average correlation among longitudinal values for a participant) different for each outcome. ICCs and their corresponding confidence intervals were estimated using a restricted-maximum-likelihood linear mixed-effects model (Diggle et al., 2002) of the outcome measure adjusted for session time, specifying random intercepts for participants. The ICC is an estimate of the average proportion of total variance attributable to true differences in the outcome measure.

Within- and between-subject correlations of median relative alpha power and other EEG-derived metrics (e.g. average P300 amplitude across all target letter events) were calculated with a fixed-effects longitudinal model (Allison, 2009) using Bland and Altman's method (Bland & Altman, 1995) for the within-subject correlation estimate and the complementary "between-effects" longitudinal estimator (used in the calculation of the fixed-effects model) for the between-subject estimate. Degrees of freedom for the within-subject correlation were estimated using the df formula noted above, and using the sample size n=5 for the between-subject correlation, in each case subtracting 3 when calculating Fisher's approximation of the z-score of the correlation (Fisher, 1920).

Longitudinal slopes over the baseline and intervention phases for each outcome measure were estimated using an ordinary linear regression of the outcome on time (in weeks), study phase (0=baseline and 1=intervention), and the time-by-phase interaction, fitted separately for each participant on each outcome. In order to account for influence of the longitudinal correlation in measured values on the measurement variance, we employed the Newey-West robust variance estimator (Newey & West, 1987) for calculating the standard errors for the slopes. Within-phase means and across-phase changes were also calculated from this model using Newey-West standard errors. To visualize the trajectory and evaluate the intervention effect on each outcome for each participant, we plotted the measurements of the outcome versus time and overlaid slopes for the baseline phase (omitting the first point) and the intervention phase. For all outcome measures, we omitted the final point from slope calculations; this point represents the follow-up session, which occurred approximately one month after the end of the intervention phase. Stata version 16.1 (StataCorp, 2019) was used for the statistical analyses listed above, and R version 3.6.1 (R core team, n.d.) was used for data management and to generate descriptive summaries.

**RESULTS**

Five participants with mild pAD enrolled in the study (See table 1 for demographic information,). All five completed the baseline phase, and two completed all baseline, intervention, and follow-up sessions. All repeated measure results for each participant are included in Table 2.

*Baseline Stability of Outcome Measures*

The stability of the outcome measures across baseline sessions is shown in Table 3. In particular, performance on WJTA-IV Sentence Reading Fluency, letter cancellation with curved foils, and AUC was reasonably stable, with CVs less than 10%. Some measures demonstrated a learning effect within the baseline phase for some participants, which can be seen in both the AUC results shown in Figure 4, outcome measure results for participants #1 and #2 (completers) in Figure 5

On the RSVP Keyboard calibration task, the five participants achieved a mean correct classification rate (AUC) across their baseline sessions of $0.72 \pm 0.03$ (mean $\pm$ SE). In the baseline phase, AUC for participants #1, #2, and #3 ranged from 0.67 to 0.80, 0.69 to 0.83, and 0.58 to 0.76, respectively. Participants #4 and #5 (both non-completers) exhibited consistently low AUC values near 0.6 across their baseline sessions (Figure 4). Illustrations of representative EEGs and ERPs taken from the 1st week of intervention for participants #1 and #2 (completers) and participant #4 (non-completer) are presented in Figure 6.

In the outcome measures tasks, letter cancellation tasks for both participants (Figure 5A, 5B, 5F, and 4G) and WJTA-IV Sentence Reading Fluency for participant #2 (Figure 5J) demonstrated baseline variability.

*Effects of Intervention*

Most outcome measures that were repeated multiple times before and during the intervention did not change significantly across phases (see Table 2, Table 4, and Figure 5). Figure 5 includes examples for which there was no change in performance with the introduction of the intervention (Figure 5D), a sustained learning effect continuing through the baseline and intervention phases (Figure 5J), a learning effect only during the baseline phase (Figure 5G), and an apparent improvement from the intervention (Figure 5A). Participant 1 demonstrated a decrease in performance in the intervention phase for the letter span forward condition but not the letter span backward condition. As the letter span backward condition is of higher complexity and may be a better index of working memory, it is believed that motivational factors or fatigue related to the difficulty of the task may have influenced this participant's performance.

Figure 4 depicts AUC results by session for each participant. AUC changed by $0.14 \pm 0.03$ (mean $\pm$ SE) across the baseline ($0.74 \pm 0.02$) and intervention ($0.89 \pm 0.01$) phases for participant #1. Participant #2's performance during the intervention phase was variable, with AUC ranging from 0.55 to 0.85 ($0.74 \pm 0.02$).

In the copy-spelling task during the intervention phase, participant #1 (mean intervention phase AUC of 0.89) correctly copied the word "HELLO" in 15/17 opportunities, compared to 0/4 opportunities in the baseline phase. Participant #2 (mean intervention phase AUC of 0.74) copied the entire word in 3/5 opportunities in the baseline phase and 7/18 opportunities in the intervention phase, and typed "HELL" correctly in an additional 5/18 intervention phase opportunities but did not complete the word. The other three participants who did not undergo the intervention had lower AUCs (as low as 0.5) for some or all of their baseline sessions and thus, not unexpectedly, were usually unable to type even the "H" correctly during baseline sessions.

The summative outcome measures (WAIS Digit Span and Discourse Comprehension Test) were given once prior to and once after the intervention. There was no significant difference between these metrics pre- and post-intervention.

*Neurofeedback, posterior alpha levels, and supplementary ERPs*

The posterior alpha power NFB was successfully implemented. The five levels of feedback were delivered with the approximate planned target frequencies. Against a targeted best-to-worst distribution of 30%, 25%, 15%, 15%, and 15%, participant #1's NFB demonstrated averages of 36.6%, 24.7%, 13.7%, 10.9%, and 14.1% across all intervention sessions. Participant #2's NFB demonstrated averages of 31.3%, 23.1%, 16.3%, 14.6%, and 14.7%. There was no clear change over time in the posterior alpha power in the two completers. There was a significant correlation between target P300 amplitude and AUC across participants (0.94; $z = 2.46$), but the within-subject relationship was not as strong (0.45; $z = 1.47$). Analogously, there was a strong correlation between N200 amplitude and AUC between participants (-0.80; $z = -1.57$) and the within-subject correlation was absent (-0.04; $z = -0.12$). Representative samples of EEG and ERP responses to targets and non-targets are presented in figure 6 in order to illustrate resting EEG including alpha activity, and typical ERP responses observed during the neurofeedback calibration task.

## DISCUSSION

In this study, two participants with mild pAD completed an intensive neurofeedback-based intervention study, with weekly home visits for over a month in the baseline phase and three visits per week for six weeks during the intervention phase. Three of five participants successfully took part in 3-5 baseline visits but were discontinued prior to or at the beginning of the intervention phase because of public health guidelines during the COVID-19 pandemic. This study demonstrated that adults with mild pAD were able to perform the BCI RSVP Keyboard calibration (letter detection) task, with one participant achieving AUC values up to 0.94 but another of the five demonstrating AUC values as low as 0.5. The participant with very low AUC appeared engaged and was able to perform the letter discrimination task, but had essentially no P300 (see Figure 6B, participant #4). There is some concern that this is at least partly related to aging that produces a decline in P300 amplitude as well as prolongation of P300 latency, and AD further increases those changes (Oken & Amen, 2010; Oken & Kaye, 1992). The task required participants to attend to 100 sequences of letters and look for a target while minimizing distractions to non-targets and artifact-inducing movements. This relies heavily on attention, a domain of cognition that is affected in early stages of AD. Additionally, participants who completed the intervention were able to correctly type words in a copy spelling task. This contributes to the BCI field by demonstrating that adults with impairments in attention, a domain of cognition, can learn to operate a BCI for communication.

Results for all five participants demonstrated that some, but not all, outcome measures were stable during the baseline phase. For instance, the WJTA-IV Sentence Reading Fluency, letter cancellation with curved foils, and AUC had CVs less than 10%. The novel letter span task with scoring similar to the WAIS Digit Span had worse baseline stability with higher CVs (19% for forward and 16% for backward). This demonstrates the stability of some measures for future studies that might be more robust to test-retest effects.

The current study demonstrated feasibility of a NFB intervention using posterior alpha power during performance of an RSVP Keyboard task. Some outcome measures showed no effect from the NFB intervention (e.g. letter span backward), and some demonstrated either a learning effect during the baseline (e.g. letter cancellation with straight foils) or a continued learning effect across the baseline and intervention periods (e.g. WJTA-IV Sentence Reading Fluency). There was an apparent intervention effect in at least one measure, with improvement on the letter cancellation task with curved foils after NFB implementation, in one of the two completers. There was no consistent change in the posterior alpha power from the NFB intervention in the two completers. Although there were no significant changes in performance across phases on most measures, it is a notable finding to see a maintenance of performance on cognitive outcome measures in the context of Alzheimer's disease, a degenerative condition impacting cognition.

This proof-of-concept study had limitations. The planned single case design was a non-blinded study with few participants. Both limit the conclusions as well as the ability to report many of the items that should be ideally reported in NFB studies, e.g., details of control group and blinding (Ros et al., 2020). Individuals with pAD who demonstrate specific language or attention impairments related to the intervention targets may benefit more from the proposed intervention than those with cognitive impairments in domains that extend beyond target intervention areas (e.g. memory, visuospatial etc.). In addition, adults with stable neurological impairments, such as chronic aphasia, may improve performance more with this innovative intervention. EEG recordings obtained in a natural environment, such as the participant's home, may be more susceptible to artifacts. However, as the purpose of this study was to assess the feasibility of an intervention not requiring participants to drive, and so it was important to conduct recordings in participants' homes in order to facilitate participant compliance, limit attrition, and to observe real-life contributors to the efficacy of intervention (e.g. typical environmental noise, dry electrode system fitting challenges, discomfort, differences in alertness etc). As with other behavioral research, future NFB studies will need to control for non-specific aspects of improvement that may have occurred with the additional social interaction and cognitive stimulation associated with data collection visits, independent of the NFB. It is possible that the baseline version of the RSVP Keyboard tasks provided stimulation that could be considered a potential intervention. The specific target for the NFB (alpha level) was not optimal for the study as; the utility of the NFB would have been supported by demonstrating changes in the EEG measure used for the NFB. Finally, the sample size was small, and homogenous in socioeconomic status, race, and education. Larger studies with control conditions and more diverse participant samples are needed to explore this question further.

## CONCLUSIONS

In conclusion, these findings constitute novel contributions to the field of AD and BCI, even though this study did not demonstrate definitive improvements in outcome measures. First, the results of this study demonstrated the feasibility of this mode of intervention in participants with mild pAD. While this project required significant programming and signal processing expertise, the use of open-source software (Memmott et al., 2021) and the availability of clinical NFB devices dramatically increases the feasibility of further research and eventual clinical application. Single case design studies with multiple assessment before and during an intervention (Gast & Spriggs, 2014; Ledford & Gast, 2018) is a potential approach for AD pilot studies. Most importantly, we have clearly demonstrated that participants with cognitive impairments such as AD can use a P300 based BCI speller for calibration and two participants were able to use the RSVP P300 speller BCI system to spell a short word ("hello") in an experimental setting. Additional research is necessary to investigate the extent and generalizability of these findings.

**DECLARATIONS**

*Availability of data and materials*

The datasets used and/or analyzed during the current study are available from the corresponding author on reasonable request.

*Competing interests*

All authors received a salary for their work from their respective institutions.

*Funding*

This research was financially supported by the grant sources R01 DC009834, DC009834-09S1, and the OHSU ADRC P30 AG066518. The sponsors had no role in study design, data collection, analysis, interpretation of data, writing of the paper, or decision for submission to publication.

*Authors' contributions*

DEM, DK, TM, BP, JW, MFO, and BO contributed to the writing of the manuscript including text, figures, and tables. DEM, DK, BO, TM, JW, and BP contributed to the design of the experiment. DEM and DK collected and processed data. DEM programmed analysis scripts and behavioral tasks outside of BciPy software. TM provided software development needed in BciPy for use as NFB intervention. JW performed statistical analyses. All authors have read and approved the manuscript, and ensure that this is the case

*Acknowledgements*

We would like to acknowledge Andy Fish and Mayling Dixon for their administrative contributions to the project.

Electroencephalography Fusion for Locked-In Syndrome. *Neurorehabilitation and Neural Repair*, *28*(4), 387–394. https://doi.org/10.1177/1545968313516867

Ordikhani-Seyedlar, M., Lebedev, M. A., Sorensen, H. B. D., & Puthusserypady, S. (2016). Neurofeedback Therapy for Enhancing Visual Attention: State-of-the-Art and Challenges. *Frontiers in Neuroscience*, *10*, 352. https://doi.org/10.3389/fnins.2016.00352

Paluch, K., Jurewicz, K., Rogala, J., Krauz, R., Szczypińska, M., Mikicin, M., Wróbel, A., & Kublik, E. (2017). Beware: Recruitment of Muscle Activity by the EEG-Neurofeedback Trainings of High Frequencies. *Frontiers in Human Neuroscience*, *11*. https://doi.org/10.3389/fnhum.2017.00119

Peirce, J., Gray, J. R., Simpson, S., MacAskill, M., Höchenberger, R., Sogo, H., Kastman, E., & Lindeløv, J. K. (2019). PsychoPy2: Experiments in behavior made easy. *Behavior Research Methods*, *51*(1), 195–203. https://doi.org/10.3758/s13428-018-01193-y

Perry, R. J., & Hodges, J. R. (1999). Attention and executive deficits in Alzheimer's disease. A critical review. *Brain: A Journal of Neurology*, *122 ( Pt 3)*, 383–404. https://doi.org/10.1093/brain/122.3.383

Pfurtscheller, G., & Lopes da Silva, F. H. (1999). Event-related EEG/MEG synchronization and desynchronization: Basic principles. *Clinical Neurophysiology: Official Journal of the International Federation of Clinical Neurophysiology*, *110*(11), 1842–1857. https://doi.org/10.1016/s1388-2457(99)00141-8

R core team. (n.d.). *R: A language and environment for statistical computing*. R Foundation for Statistical Computing. http://www.R-project.org/
35

**CONSORTIUM MEMBERSHIP**

The Consortium for Accessible Multimodal Brain-Body Interfaces (CAMBI) was comprised of the following members during the time of this study (excluding authors who are affiliated with CAMBI): Deniz Erdogmus[7], David Smith[8], Steven Bedrick[9], Brandon Eddy[10], Michelle Kinsella[1], Matthew Lawhead[11], Aziz Kocanaogullari[7], and Shiran Dudy[9].

[1] Institute on Disability and Development, Department of Pediatrics, Oregon Health & Science University, Portland, OR, USA

[7] Department of Electrical and Computer Engineering, College of Engineering, Northeastern University, Boston MA

[8] Khoury College of Computer Sciences, Northeastern University, Boston, MA, USA

[9] Department of Medical Informatics and Clinical Epidemiology, Oregon Health & Science University, Portland, OR, USA

[10] Department of Speech & Hearing Sciences, Portland State University, Portland, OR, USA

[11] Oregon Clinical and Translational Research Institute, Oregon Health & Science University, Portland, OR, USA

**FIGURES**

Figure 1. Study Activities Table. This infographic depicts the assessments conducted at each visit type and the frequency of their measure across phrases of the study. The dashed lines in the baseline phase represent the variability in number of baseline visits that varied between 4 to 7 visits based on stability of baseline performance from week-to-week.

| | Study Entry | Baseline | | | | | | | Intervention | | | | | | Follow-Up |
|---|---|---|---|---|---|---|---|---|---|---|---|---|---|---|---|
| Week in phase: | 1 | 1 | 2 | 3 | 4 | 5 | 6 | 7 | 1 | 2 | 3 | 4 | 5 | 6 | 1 |
| Consenting | X | | | | | | | | | | | | | | |
| DCT (raw score) | X | | | | | | | | | | | | | | |
| WAIS-IV digit span (raw score) | X | | | | | | | | | | | | | | |
| Practice RSVP task | X | X | X | X | X | X | X | X | X X X | X X X | X X X | X X X | X X X | X X X | X |
| RSVP calibration – no NFB | | X | X | X | X | X | X | X | | | | | | | X |
| RSVP calibration – NFB | | | | | | | | | X X X | X X X | X X X | X X X | X X X | X X X | |
| RSVP copy-phrase | | X | X | X | X | X | X | X | X X X | X X X | X X X | X X X | X X X | X X X | X |
| Letter cancellation | | X | X | X | X | X | X | X | X | X | X | X | X | X | X |
| Letter span task | | X | X | X | X | X | X | X | X | X | X | X | X | X | X |
| WJTA sentence reading fluency | | X | X | X | X | X | X | X | X | X | X | X | X | X | X |

Figure 2. Schematic of the RSVP task and the neurofeedback to the participant. Following presentation of the target letter there is a sequence of 10 letters presented following a red cross-hair warning signal. Neurofeedback is based on individualized alpha Power Spectral Density (PSD) percentiles at a pre-specified occipitoparietal electrode. Colored boxes range from upper 15th percentile alpha power in red, 15th - 30th percentile, 30th - 45th, 45th to 70th, and the least alpha power in green from the 70th - 100th percentiles. The current feedback the participant sees is highlighted with the white edges around one colored rectangle

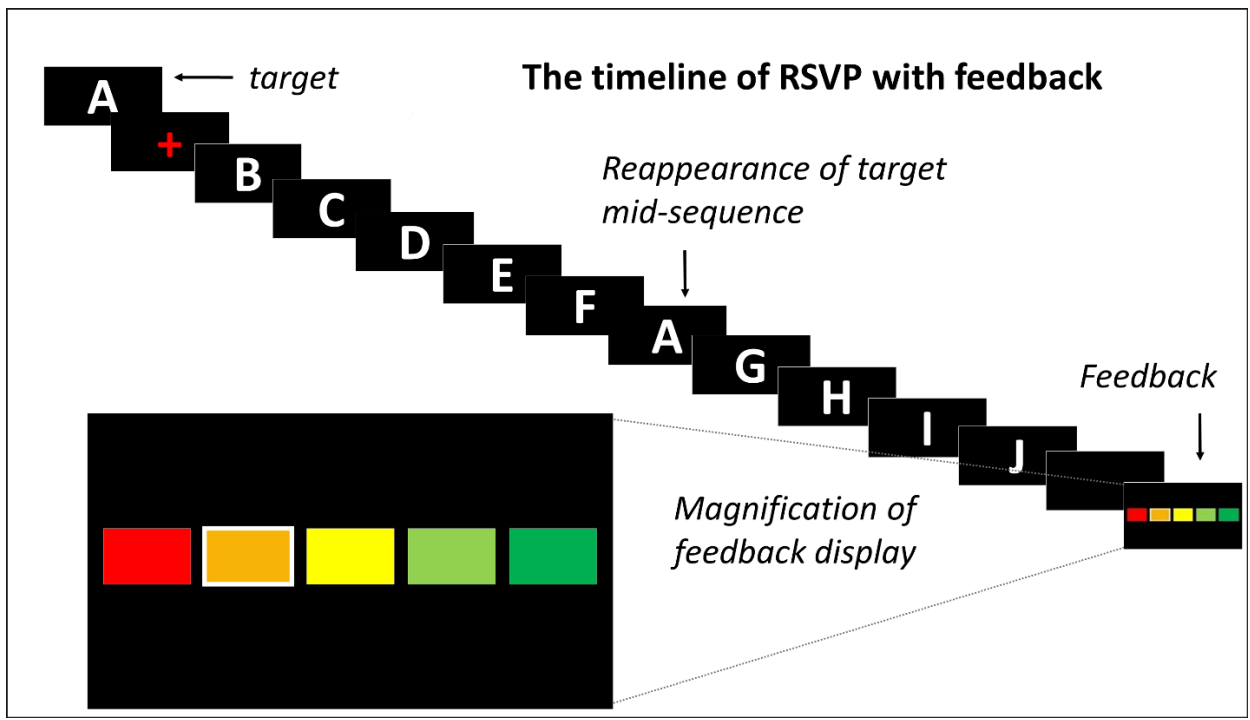

Figure 3. EEG frequency data from 5 of 8 participants who made errors during pilot task. Spectra are from site Oz, averaged across 2.5 s epochs during sequences where participants either made or did not make recognition errors. Besides the increased alpha power during error trials, note the steady-state visual evoked potential aligned with presentation rate of 4 Hz, which did not change based on error status. The three participants not included in this average because they did not make errors had even lower levels of alpha than the "correct" condition illustrated in the figure.

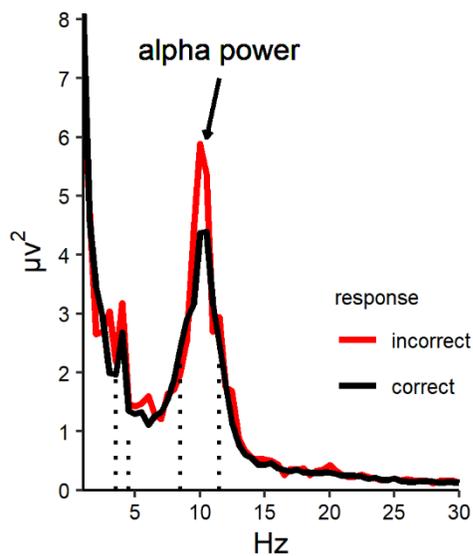

Figure 4. AUC scores for five participants. BCI sessions were conducted once per week in the baseline phase, and three times per week in the intervention phase. Curves showing the trajectory of AUC for each participant across the study duration were calculated using a running-mean lowess smoother with tricube weighting and default bandwidth. Three participants demonstrated a learning effect during the baseline. One participant (#1) continued to improve during the intervention, with AUC values rising to 0.90. Another participant with a particularly low AUC (#4), in some sessions about 0.5, had no discernible N200-P300 in the averaged ERP waveforms.

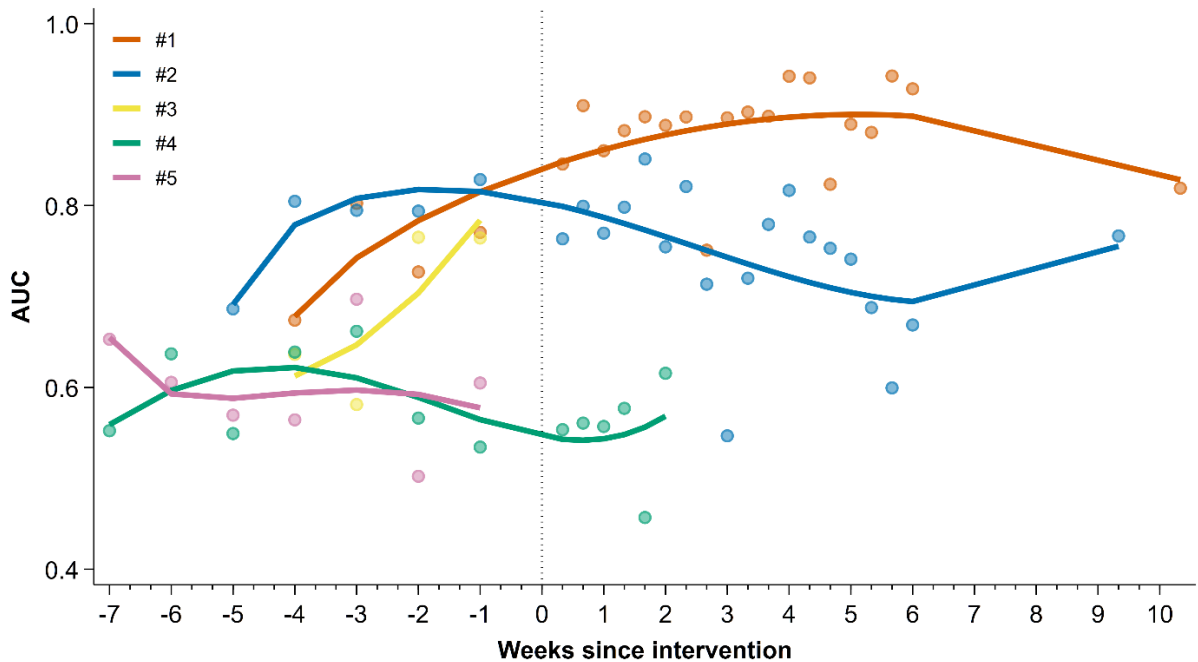

Figure 5. Full data from repeated measures tasks administered in baseline sessions, intervention sessions, and follow-up session for the two participants who completed the study. In general, the WJTA IV Sentence Reading Fluency, Letter Cancellation with curved foils, and AUC (Figure 2) were the most reliable. Note examples of outcome measures in for which there was: no change in the intervention (Letter Span Backward) (d and i); a sustained learning effect continuing through the baseline and then through the intervention periods (Sentence Reading Fluency) (j); a learning effect only during the baseline (Letter cancellation with straight foils) (b and g); an apparent improvement from the intervention (Letter Cancellation with curved foils) (a), and a decrease in performance in treatment phase (Letter Span forward) (c).

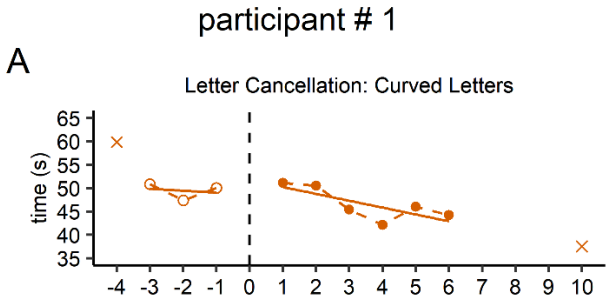
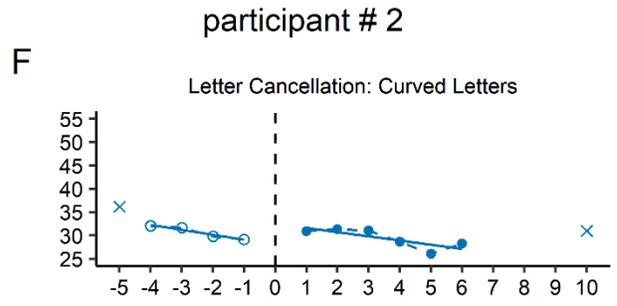
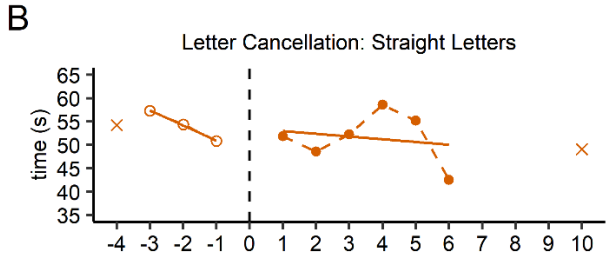
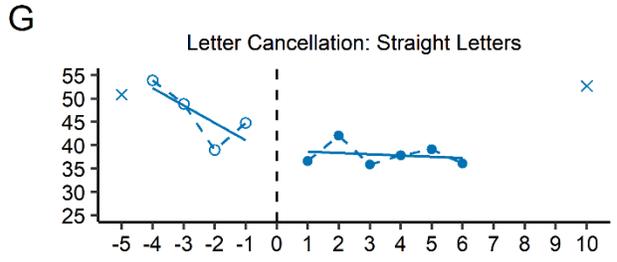
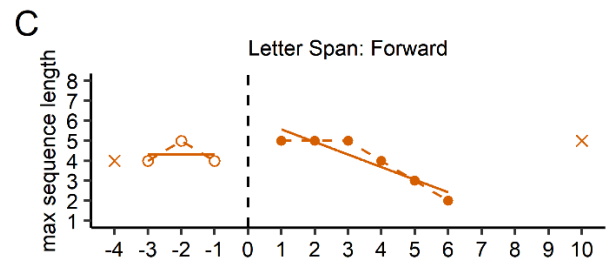
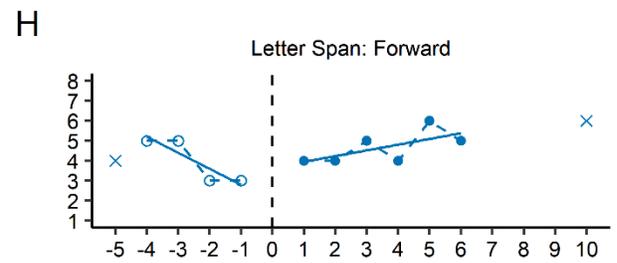
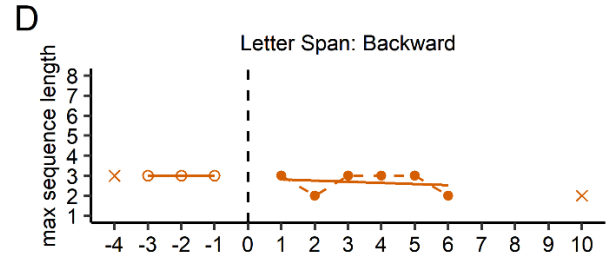
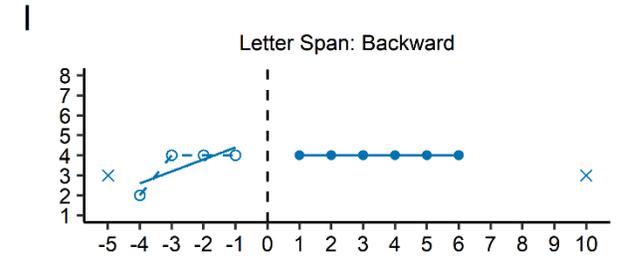
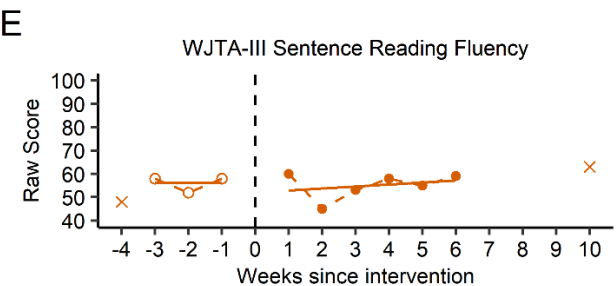
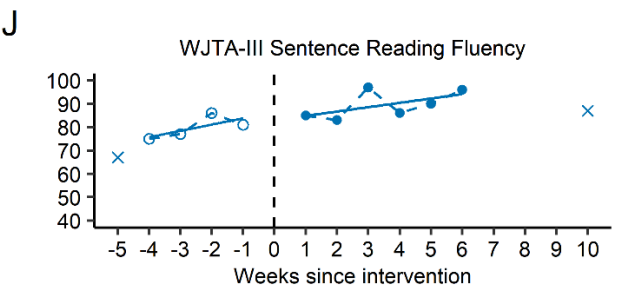

Figure 6. (A) 5-seconds of representative EEG data taken from week #1 of intervention. The presented windows each include one full sequence of 10 letters (1 target; 9 non-targets) during the RSVP neurofeedback calibration task. Parieto-occipital alpha is clearly visible at sites P4 and Pz prior to fixation. Participant #1 demonstrates neck EMG contamination at posterior sites Oz, PO7, and PO8. (B) Demonstrative ERP averages of target and non-target responses at Pz, derived from week #1 of intervention (3 sessions; neurofeedback calibration). Participant #1 shows EMG contamination, but a large N2/P3 response. Participant #2 exhibits alpha signal but also a small yet clear ERP response to the target. Participant #4 demonstrates no visible target-related response.

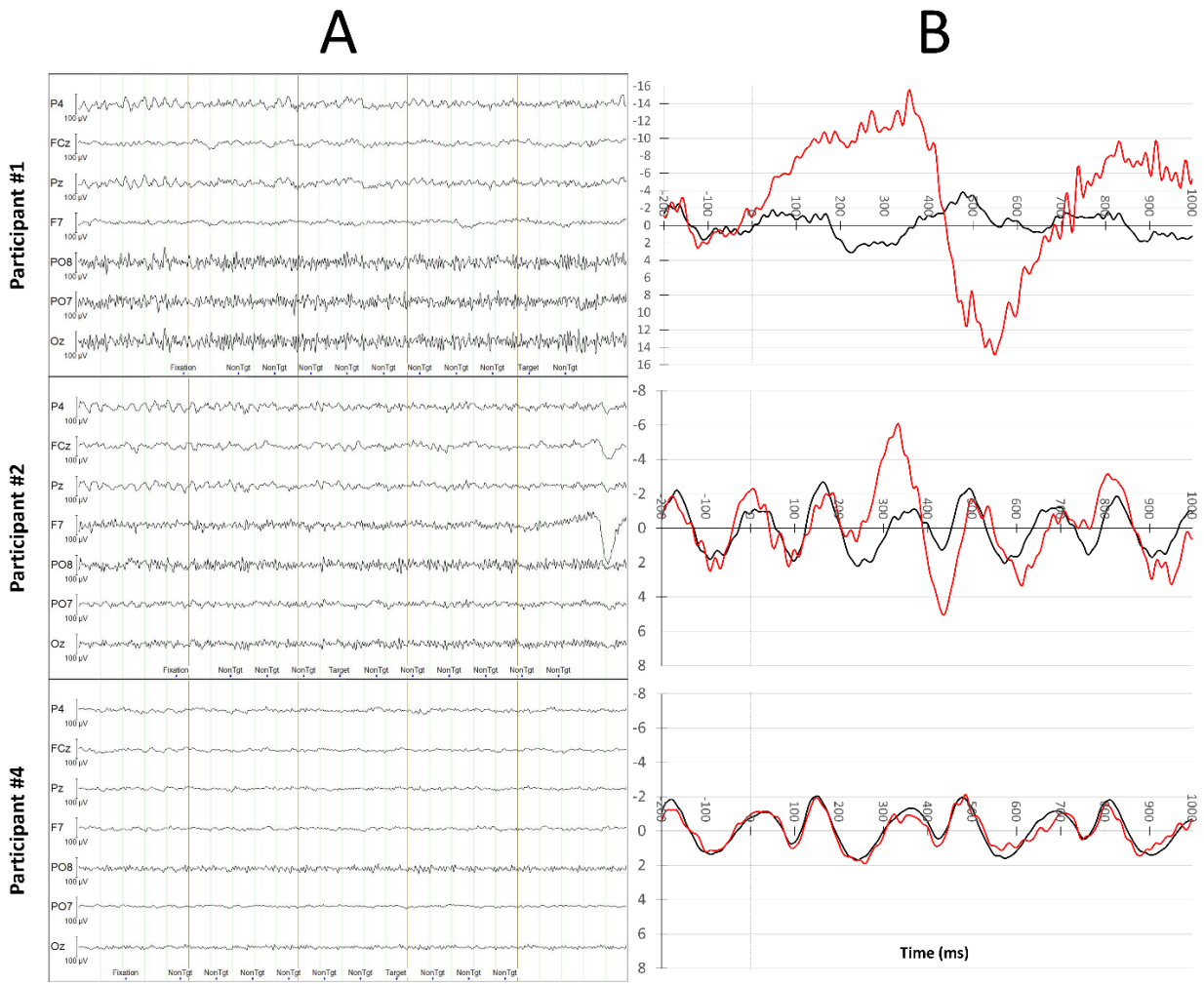

# TABLES

Table 1. Participant demographics. All participants were white and non-Hispanic

| Participant | Age | Years of Education | MoCA | # Baseline Sessions | # Intervention Sessions |
|---|---|---|---|---|---|
| **#1** | 79 | 16 | 21 | 4 | 18 |
| **#2** | 66 | 17 | 29 | 5 | 18 |
| **#3** | 53 | 18 | 22 | 4 | --- |
| **#4** | 72 | 18 | 24 | 7 | 6 |
| **#5** | 76 | 13 | 19 | 7 | --- |

Table 2. This table shows the raw score performance on all repeated measures for both baseline and intervention phases for each participant. The number of sessions is included for reference as only two participants (participant #1 and #2) completed the study in its entirety. The asterisk (*) is used to demonstrate which participants performed below expectations as defined as deviation from the expected raw score based on comparison sample of same-aged peers without known impairments. For the letter cancellation task, no measure of variance or range of performance was included in the descriptive statistics for any participant group. It is therefore not reliable to make a judgement on participants' performance without a range of performance. Averages are included as a reference of performance. Same-aged peers without AD completed the letter cancellation task in an average of 25 seconds for curved letters and 32 seconds for straight letters (Baddeley et al., 2001). Participants with AD completed the letter cancellation task in an average of 35 seconds for curved letters and 49 seconds for straight letters (Baddeley et al., 2001). For the WJTA-IV sentence reading fluency subtest, average performance as defined as receiving a standard score between 90 and 110 is equivalent to a raw score of 58-85 for ages 50-59, 54-80 for ages 60-69, and 49-74 for ages 70-79 (normative sample used for comparison). There was no same-aged peer group in literature for letter span task to make comparisons of performance.

| Participant | # of Sessions | Measure | Baseline (M ± SD) | Intervention (M ± SD) |
|---|---|---|---|---|
| # 1 | Baseline: 4 Intervention: 18 | Letter Cancellation Curved Letters Straight Letters | 52.03 ± 5.42 54.19 ± 2.65 | 46.56 ±3.55 51.51 ± 5.53 |
| | | Letter Span Forward Backward | 4.25 ± 0.5 3 ± 0 | 4 ± 1.26 2.67 ± 0.52 |
| | | WJTA-IV SRF | 54 ± 4.9 | 55 ± 5.55 |
| # 2 | Baseline: 5 Intervention: 18 | Letter Cancellation Curved Letters Straight Letters | 31.76 ± 2.74 47.5 ± 5.79 | 29.35 ± 2.07 37.93 ± 2.38 |
| | | Letter Span Forward Backward | 4 ± 1 3.4 ± 0.89 | 4.6 ± 0.81 4 ± 0 |
| | | WJTA-IV SRF | 77.2 ± 7.08 | 89.5 ± 5.89 |
| # 3 | Baseline: 4 Intervention: 0 : | Letter Cancellation Curved Letters Straight Letters | 47.55 ± 16.90 57.21 ± 14.24 | -- |
| | | Letter Span Forward Backward | 5.25 ± 0.5 4 ± 0 | |
| | | WJTA-IV SRF | 61.25 ± 8.18* | |
| # 4 | Baseline: 7 Intervention: 6 | Letter Cancellation Curved Letters Straight Letters | 35.8 ± 2.48 40.01 ± 3.31 | 33.9 ± 1.53 39.88 ± 3.53 |
| | | Letter Span Forward Backward | 4.86 ± 0.9 4.43 ± 0.53 | 6 ± 0 5 ± 0 |
| | | WJTA-IV SRF | 49.43 ± 3.26* | 48 ± 7.07* |
| # 5 | Baseline: 7 Intervention: 0 | Letter Cancellation Curved Letters Straight Letters | 39.67 ± 2.35 50.44 ± 4.54 | -- |
| | | Letter Span Forward Backward | 3.71 ± 0.49 2.85 ± 0.38 | |
| | | WJTA-IV SRF | 56.2 ± 3.27 | |

Table 3. Within-person baseline outcome measure stability (excluding the first baseline session for each participant), averaged across the five participants. Coefficients of Variation (CV) greater than 10% indicate poor stability. Given the number of participants, the wide range of ICC estimates is difficult to interpret but consistent with the CVs. ICCs are very high for WJ-Sentence Reading Fluency and Letter Cancellation Time: curved-letter foils

| Measure | CV ± SE (%) | ICC [95% CI] |
|---|---|---|
| AUC | 9.23 ± 2.57 | 0.68 [0.28, 0.92] |
| Median Relative Alpha PSD | 24.90 ± 6.68 | ≈0 |
| WJTA-IV: Sentence Reading Fluency Raw Score | 6.50 ± 1.98 | 0.90 [0.65, 0.98] |
| Maximum Letter Span: Forward | 19.01 ± 4.11 | 0.22 [0.02, 0.80] |
| Maximum Letter Span: Backward | 16.04 ± 4.20 | 0.57 [0.18, 0.89] |
| Letter Cancellation (sec): Curved | 5.19 ± 1.59 | 0.92 [0.71, 0.98] |
| Letter Cancellation (sec): Straight | 13.38 ± 3.17 | 0.37 [0.06, 0.83] |

Table 4. Longitudinal slopes over the baseline and intervention periods for each outcome measure. Slope estimates were generated using an ordinary linear regression of the outcome on time, and standard errors employed the Newey-West robust variance estimator. Note that a negative sign indicates improvement for some measures (both letter cancellation tests), but otherwise a positive sign represents improvement.

| Measure | Slope ± SE | Participant #1 | Participant #2 |
|---|---|---|---|
| **SRF Raw Score** | Baseline | 0.0 ± 4.1 | 2.7 ± 2.2 |
|  | Intervention | 0.9 ± 1.4 | 1.9 ± 1.2 |
| **Max Letter Span: Forward** | Baseline | 0.0 ± 0.4 | -0.8 ± 0.3 |
|  | Intervention | -0.6 ± 0.1 | 0.3 ± 0.2 |
| **Max Letter Span: Backward** | Baseline | 0.0 ± 0.4 | 0.6 ± 0.2 |
|  | Intervention | -0.1 ± 0.1 | 0.0 ± 0.1 |
| **Letter Cancellation (sec): Curved** | Baseline | -0.4 ± 1.8 | -1.1 ± 0.5 |
|  | Intervention | -1.5 ± 0.6 | -0.9 ± 0.3 |
| **Letter Cancellation (sec): Straight** | Baseline | -3.2 ± 3.8 | -3.7 ± 1.6 |
|  | Intervention | -0.6 ± 1.3 | -0.3 ± 0.9 |